\documentclass[aps,prmaterials,reprint,superscriptaddress,amsmath,amssymb,floatfix]{revtex4-2}
\usepackage{graphicx}
\usepackage{bm}
\usepackage{xcolor}
\usepackage[hidelinks]{hyperref}
\newcommand{\Ainv}{\AA$^{-1}$}
\newcommand{\uC}{\textmu C\,cm$^{-2}$}
\newcommand{\NOMBA}{4-NO$_2$MBA$_2$PbBr$_4\cdot$H$_2$O}

\begin{document}

\title{Polarization-locked chiral persistent spin textures in two-dimensional lead-halide perovskites}

\author{Palanichamy Gayathri}
\email[Corresponding author: ]{gayathri.palanichamy@physics.uu.se}
\author{Biplab Sanyal}
\author{Jonas Fransson}
\email[Corresponding author: ]{jonas.fransson@physics.uu.se}
\affiliation{Department of Physics and Astronomy, Uppsala University, Box-516, 75237, Uppsala, Sweden}

\date{\today}

\begin{abstract}
Chiral spacer cations crystallize two-dimensional hybrid perovskites in the polar Sohncke group $P2_1$, which places the
polar axis in the inorganic layer. We analyze the spin-orbit field that this geometry allows and test it against first-principles spin
textures of ten polar chiral lead-halide perovskites, their mirror enantiomers and four
halide-substituted analogues. The point group $C_2$ with the twofold axis in the layer forbids the helical Rashba field and
allows a linear term $\gamma_1k_x\sigma_z$, the Rashba interaction of the in-plane polar axis, together with
chirality-related Dresselhaus and radial terms. Thirty-four two-band $\mathbf k\cdot\mathbf p$ models were fitted to each
band pair and ranked by the agreement of their spin directions with the calculation rather than by the band energies,
which cannot distinguish them. In all strongly polar compounds, the conduction band is described by the persistent term
alone, with $\gamma_1=1.09$--1.98\,eV\,\AA{} and a spin agreement of at least 0.98, whereas the best Rashba or Dresselhaus
model reaches 0.20. The R and S enantiomers of each compound have opposite polarizations, and therefore opposite $\gamma_1$ and
opposite persistent spin textures of equal magnitude, as required by the mirror that relates them; this holds in all ten
compounds, so the choice of the enantiomer selects the spin direction. The magnitude of $\gamma_1$ correlates with the disparity of the
in-plane Pb-X-Pb angles ($r=0.95$). With electron masses of 0.26--0.43\,$m_e$ along the spin-splitting directions, the
associated persistent spin helices have wavelengths of 3.4--8.2\,nm.
\end{abstract}

\maketitle

\section{Introduction}
Spin-orbit coupling (SOC) in crystals without inversion symmetry lifts the spin degeneracy of the bands and locks the spin
to the crystal momentum. Near a band edge this is described by a momentum-dependent effective magnetic field, the
spin-orbit field, whose form is fixed by the point group; the helical Rashba field of structures with an out-of-plane
polar axis and the Dresselhaus field of non-polar acentric crystals are the best-known
examples~\cite{bychkov1984,dresselhaus1955,manchon2015}. In halide perovskites these effects are large, because
their conduction band is formed by Pb-6$p$ states with a large atomic SOC~\cite{even2013,kepenekian2017}, and large
Rashba-type splittings have been predicted and observed~\cite{stroppa2014,zhai2017}. In two-dimensional (2D) hybrid
perovskites, the organic spacer cations also control the symmetry and the distortions of the inorganic layers, and with
them the form and the size of the spin-orbit field~\cite{saparov2016,katan2019,blancon2020,jana2021}.

If the spin-orbit field points along the same direction for all momenta, the spin component along
this direction is conserved, the spin texture is persistent, and spin relaxation by the D'yakonov-Perel' mechanism is
suppressed; an injected spin precesses into a persistent spin helix (PSH) whose wavelength is set by the strength of the
coupling~\cite{dyakonov1972,schliemann2003,bernevig2006,schliemann2017}. In III-V quantum wells such a persistent spin
texture (PST) requires the Rashba and Dresselhaus couplings to be tuned to equal strength~\cite{koralek2009}. Tao and
Tsymbal showed that it can instead be enforced by the point group, without any tuning, when the polar axis lies in the
plane of a 2D system~\cite{tao2018}. Symmetry-protected PSTs of this kind have since been reported for a layered
ferroelectric oxide~\cite{autieri2019}, group-IV monochalcogenide monolayers~\cite{absor2019}, ferroelectric SnTe
films~\cite{lee2020} and, among 2D hybrid perovskites, achiral polar crystals with an in-plane polarization~\cite{jia2020,zhang2022}.

Chiral 2D perovskites, in which chiral spacer cations transfer their handedness to the inorganic
framework~\cite{long2020,jana2020}, belong to the Sohncke space groups, which contain only proper symmetry
operations~\cite{fecher2022}. Their SOC has attracted interest in connection with spin-dependent transport and
chirality-induced spin selectivity~\cite{lu2019}, and their spin splitting has been correlated with the distortions of the
inorganic layer~\cite{jana2020,jana2021}. Many of these compounds crystallize in the polar Sohncke group $P2_1$, in which
the $2_1$ screw axis, and therefore the polar axis, lies in the inorganic layer. The polar distortion arises from the
coupling of octahedral tilts with displacements of the spacer cations, a mechanism related to hybrid improper
ferroelectricity~\cite{benedek2011,gayathri2024,gayathri2025,ghosh2026,zheng2024}.

Two questions remain open for these crystals. First, the point group of $P2_1$, $C_2$, differs from the $C_{2v}$ group of the
achiral polar systems studied so far: without mirror planes it allows, besides the persistent term, Dresselhaus and radial
terms associated with the handedness. Whether the persistent term dominates has to be established, and band energies
alone cannot do this, because linear Rashba, Dresselhaus, radial and persistent fields produce identical splittings and
differ only in the spin directions. Second, chiral crystals exist as two enantiomers, and it is
not known how the persistent spin textures of the R and S crystals are related and what sets their sign.

Here we answer both questions for ten polar chiral lead-halide perovskites, their mirror
enantiomers and four halide-substituted analogues. We first write down the spin-orbit field allowed by $C_2$ with the
method of invariants, solve the resulting two-band model and derive how its coefficients transform between enantiomers and
polarization domains (Sec.~\ref{sec:theory}). We then fit 34 two-band
$\mathbf k\cdot\mathbf p$ models to density-functional spin textures and select the model by the agreement of the spin
directions (Sec.~\ref{sec:methods}). In all strongly polar compounds the conduction band is described by the persistent term
alone; the R and S enantiomers have opposite polarizations and opposite $\gamma_1$, so that they carry opposite
persistent spin textures, and the magnitude of $\gamma_1$ is set by the in-plane distortion of the Pb-X network
(Sec.~\ref{sec:results}).

\section{Symmetry analysis and model Hamiltonian}\label{sec:theory}

\begin{figure*}[t]
\includegraphics[width=\textwidth]{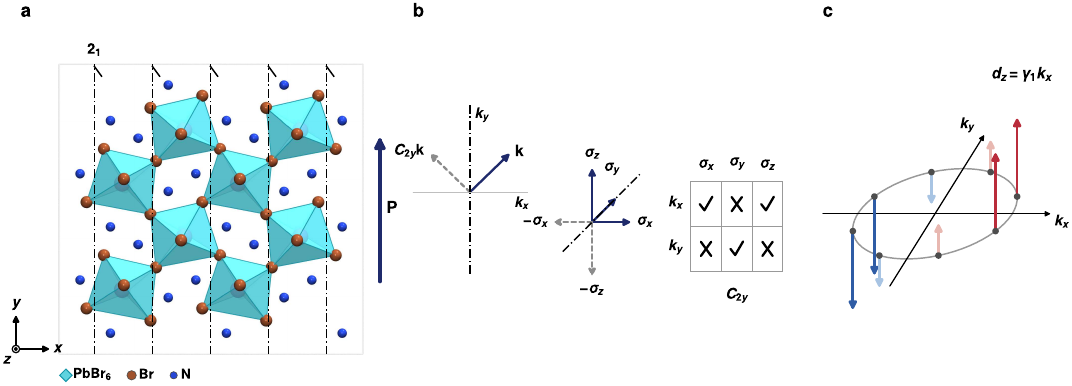}
\caption{\label{fig:frame}Frame and symmetry of the spin-orbit field. (a) One PbBr$_4$ layer of S-\NOMBA{} ($2\times2$ cells)
viewed along the layer normal $z$; the organic cations are omitted except the ammonium N. Dash-dotted lines: $2_1$ screw axes
along $y\parallel b$; navy arrow: in-plane polarization $\mathbf P$ along $+y$; $x=\hat{\mathbf y}\times\hat{\mathbf z}$.
(b) Method of invariants for the twofold rotation $C_{2y}$. Left: $C_{2y}$ maps the wave vector, a polar vector, from
$(k_x,k_y)$ to $(-k_x,k_y)$ (dash-dotted line: rotation axis). Middle: the spin is an axial vector, so $C_{2y}$ reverses
$\sigma_x$ and $\sigma_z$ (gray, dashed) and keeps $\sigma_y$. Right: the resulting sign of each product $k_i\sigma_j$; only
$k_x\sigma_x$, $k_y\sigma_y$ and $k_x\sigma_z$ are invariant ($\checkmark$) and may enter $H$, the other three are forbidden
($\times$) (Table~\ref{tab:invariants}). (c) The allowed out-of-plane field $\mathbf d=\gamma_1k_x\hat{\mathbf z}$ on a
circle around $\Gamma$ (oblique view of the $k_x$-$k_y$ plane), colored by its sign: normal to the layer at every
$\mathbf k$, reversing with $k_x$ and vanishing along $k_y$.}
\end{figure*}

\subsection{Two-band Hamiltonian near a band edge}
We use a Cartesian frame with $y$ along the $2_1$ axis, which is the polar axis $b$, $z$ along the layer normal and
$x=\hat{\mathbf y}\times\hat{\mathbf z}$ in the layer [Fig.~\ref{fig:frame}(a)]. Each band of a crystal without inversion
symmetry is spin split, but time reversal keeps the two spin branches degenerate at $\Gamma$, where they form a Kramers
pair. Near $\Gamma$ such a pair is described by the $2\times2$ Hamiltonian
\begin{equation}
H(\mathbf k)=E_0+\frac{\hbar^2k^2}{2m^*}+\mathbf d(\mathbf k)\cdot\bm\sigma,\qquad \mathbf k=(k_x,k_y).
\label{eq:H}
\end{equation}
Here $E_0$ is the energy of the Kramers point at $\Gamma$, $m^*$ the in-plane effective mass, $k^2=k_x^2+k_y^2$, and
$\bm\sigma=(\sigma_x,\sigma_y,\sigma_z)$ are the Pauli matrices acting on the spin. The first two terms describe a
spin-degenerate parabolic band. The last term is the spin-orbit coupling: $\mathbf d(\mathbf k)=(d_x,d_y,d_z)$ is the
spin-orbit field, an effective, momentum-dependent magnetic field in energy units that acts on the spin. Only in-plane
wave vectors appear because the layers are separated by the organic cations, so that the band-edge states are nearly
flat along the stacking direction (Sec.~\ref{sec:mass}).

Time reversal $\mathcal T$ reverses both the wave vector and the spin, $\mathbf k\to-\mathbf k$ and
$\bm\sigma\to-\bm\sigma$. Invariance of $H$ under $\mathcal T$ therefore requires $\mathbf d(-\mathbf k)=-\mathbf
d(\mathbf k)$: the spin-orbit field is odd in $\mathbf k$, and its expansion contains only odd powers of the wave vector.
To lowest order, each component of $\mathbf d$ is a linear combination of $k_x$ and $k_y$, so that $\mathbf
d\cdot\bm\sigma$ is a sum of the six products $k_i\sigma_j$ ($i=x,y$; $j=x,y,z$) with real coefficients. The crystal
symmetry decides which of them may appear.

\subsection{Linear invariants of the point group $C_2$}
The point group of $P2_1$ is $C_2=\{E,C_{2y}\}$, where $C_{2y}$ is a rotation by $\pi$ about $y$; the fractional
translation of the screw axis does not act on $\mathbf k$ near $\Gamma$. A product $k_i\sigma_j$ may enter $H$ only if it
is invariant under $C_{2y}$ (method of invariants~\cite{tao2018,ganguli2025}). The rotation acts on the wave vector, a
polar vector, as $(k_x,k_y)\to(-k_x,k_y)$. The spin is an axial vector; an axial vector transforms like a polar vector
under rotations, so $(\sigma_x,\sigma_y,\sigma_z)\to(-\sigma_x,\sigma_y,-\sigma_z)$ [Fig.~\ref{fig:frame}(b)]. Each
product acquires the product of the two signs, listed in Table~\ref{tab:invariants}. Three products are invariant,
$k_x\sigma_x$, $k_y\sigma_y$ and $k_x\sigma_z$, and the most general linear field is
\begin{equation}
\mathbf d\cdot\bm\sigma=\gamma_1k_x\sigma_z-\beta_1(k_x\sigma_x-k_y\sigma_y)+w_1(k_x\sigma_x+k_y\sigma_y).
\label{eq:d}
\end{equation}
Here the two in-plane invariants $k_x\sigma_x$ and $k_y\sigma_y$ are written as their antisymmetric and symmetric
combinations, which correspond to the Dresselhaus and radial textures. All three coefficients have units of eV\,\AA. The
coefficient $\gamma_1$ gives a field along the layer normal $z$ whose sign changes with $k_x$; this is the persistent term
discussed below. The coefficient $\beta_1$ is the linear Dresselhaus coefficient, an in-plane field that winds around $\Gamma$
in the sense opposite to $\mathbf k$. The coefficient $w_1$ describes the radial (Weyl) field, in which the spin is parallel or
antiparallel to $\mathbf k$; the isotropic form $\mathbf k\cdot\bm\sigma$ is a pseudoscalar and occurs only in chiral point
groups. In components, $d_x=(w_1-\beta_1)k_x$, $d_y=(w_1+\beta_1)k_y$ and $d_z=\gamma_1k_x$. The helical Rashba term
$\alpha_1(k_y\sigma_x-k_x\sigma_y)$, which describes 2D perovskites with an out-of-plane polar axis, contains the forbidden
products $k_y\sigma_x$ and $k_x\sigma_y$ and is absent, and so is $k_y\sigma_z$.

The persistent term is the Rashba interaction of the in-plane polar axis. For a polar direction $\hat{\mathbf e}$, the
Rashba field is $\alpha\,\hat{\mathbf e}\cdot(\bm\sigma\times\mathbf k)$; for $\hat{\mathbf e}=\hat{\mathbf z}$ this is
the familiar helical term, and for $\hat{\mathbf e}=\hat{\mathbf y}$
\begin{equation}
\hat{\mathbf y}\cdot(\bm\sigma\times\mathbf k)=\sigma_zk_x-\sigma_xk_z\;\xrightarrow{\;k_z=0\;}\;k_x\sigma_z .
\label{eq:rashba}
\end{equation}
An in-plane polarization thus turns the helical Rashba field into a field normal to the layer, with $\gamma_1$ playing the
role of the Rashba parameter [Fig.~\ref{fig:frame}(c)].

\begin{table}[b]
\caption{\label{tab:invariants}Behavior of the linear products $k_i\sigma_j$ under the twofold rotation $C_{2y}$ of the
point group $C_2$: sign of $k_i$, sign of $\sigma_j$ and sign of the product. A product is allowed in $H$ ($\checkmark$)
only if it is invariant (sign $+$). All products are even under time reversal.}
\begin{ruledtabular}
\begin{tabular}{lcccc}
Product & $k_i$ & $\sigma_j$ & $k_i\sigma_j$ & Allowed\\
\hline
$k_x\sigma_x$ & $-$ & $-$ & $+$ & $\checkmark$\\
$k_x\sigma_y$ & $-$ & $+$ & $-$ & $\times$\\
$k_x\sigma_z$ & $-$ & $-$ & $+$ & $\checkmark$\\
$k_y\sigma_x$ & $+$ & $-$ & $-$ & $\times$\\
$k_y\sigma_y$ & $+$ & $+$ & $+$ & $\checkmark$\\
$k_y\sigma_z$ & $+$ & $-$ & $-$ & $\times$\\
\end{tabular}
\end{ruledtabular}
\end{table}

\subsection{Other Sohncke groups and cubic terms}
The same procedure applied to the two neighboring Sohncke groups of the family gives the selection rules of
Table~\ref{tab:sym}. The non-polar group $P2_12_12_1$, to which several chiral 2D perovskites belong, has the point group
$D_2$ with three perpendicular twofold axes. The additional rotation $C_{2x}$, $(k_x,k_y)\to(k_x,-k_y)$ and
$(\sigma_x,\sigma_y,\sigma_z)\to(\sigma_x,-\sigma_y,-\sigma_z)$, reverses $k_x\sigma_z$, so the persistent term is
forbidden and only the in-plane Dresselhaus and radial terms remain. In the triclinic group $P1$ ($C_1$) every term is
allowed, including $k_y\sigma_z$; the polar axis is not fixed by symmetry, and the persistent field may contain both
$\gamma_1k_x\sigma_z$ and $\gamma_2k_y\sigma_z$. At third order in $\mathbf k$, the cubic Dresselhaus term
$-4\beta_3k_xk_y(k_y\sigma_x-k_x\sigma_y)$ is allowed in $C_2$, whereas the cubic Rashba term
$\alpha_3[(3k_x^2-k_y^2)k_y\sigma_x-(k_x^2-3k_y^2)k_x\sigma_y]$ is forbidden. Both were included in the fits to test the
symmetry analysis independently.

\subsection{Bands and spin textures}
\begin{figure*}[t]
\includegraphics[width=\textwidth]{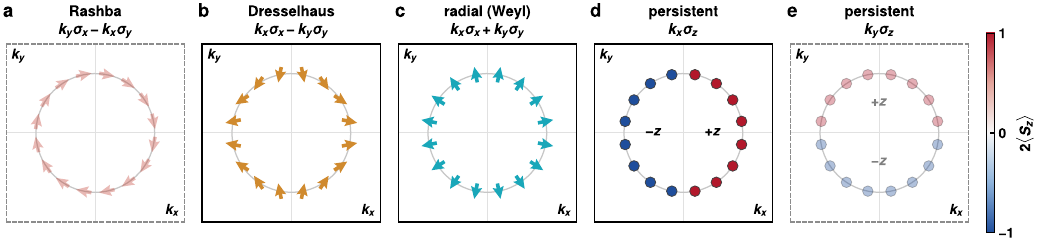}
\caption{\label{fig:terms}Spin direction of the upper branch, $\langle\mathbf S\rangle=(\hbar/2)\,\mathbf d/|\mathbf d|$, on a
circle $|\mathbf k|=$ const around $\Gamma$ for each linear spin-orbit term alone [Eq.~(\ref{eq:exact})]; the lower branch has
the opposite spin. (a) Rashba, $k_y\sigma_x-k_x\sigma_y$: tangential spins with winding $+1$. (b) Dresselhaus,
$k_x\sigma_x-k_y\sigma_y$: winding $-1$. (c) Radial (Weyl), $k_x\sigma_x+k_y\sigma_y$: spin parallel to $\mathbf k$.
(d) Persistent, $k_x\sigma_z$: spin along $+z$ for $k_x>0$ and $-z$ for $k_x<0$. (e) Persistent, $k_y\sigma_z$: spin set by
the sign of $k_y$. In-plane spins are drawn as arrows in the color of the term, out-of-plane spins as circles colored by
$2\langle S_z\rangle$. Solid frames: terms allowed in $P2_1$ (point group $C_2$ with the twofold axis along $y$), panels
(b)--(d); dashed frames and faded symbols: forbidden terms, panels (a) and (e).}
\end{figure*}

For any combination of terms, Eq.~(\ref{eq:H}) is a $2\times2$ problem that is solved exactly. Writing $\mathbf
d=|\mathbf d|\hat{\mathbf d}$, the operator $\hat{\mathbf d}\cdot\bm\sigma$ has the eigenvalues $\pm1$, so that
\begin{equation}
E_\pm(\mathbf k)=E_0+\frac{\hbar^2k^2}{2m^*}\pm|\mathbf d(\mathbf k)|,\qquad
\langle\mathbf S\rangle_\pm=\pm\frac{\hbar}{2}\frac{\mathbf d(\mathbf k)}{|\mathbf d(\mathbf k)|}.
\label{eq:exact}
\end{equation}
The two branches $E_\pm$ are split by $2|\mathbf d(\mathbf k)|$, and the spin $\langle\mathbf S\rangle_\pm$ of each branch
is fully polarized along $\pm\mathbf d$. The energies depend only on the length of the spin-orbit field, the spins only on
its direction. The linear Rashba, Dresselhaus, radial and persistent terms all give $|\mathbf d|=\alpha k$ with $k$ along
the appropriate direction, so their bands are identical (up to the direction of the splitting for the persistent term),
and only the spin textures distinguish them~\cite{ganguli2025} (Fig.~\ref{fig:terms}): tangential with winding $+1$ (Rashba), tangential with
winding $-1$ (Dresselhaus), radial (Weyl), and out of the layer (persistent). A model can therefore reproduce the
calculated band energies and still give the wrong spins, and for this reason we select the models by the spin directions.

\begin{table}[b]
\caption{\label{tab:sym}Spin-orbit terms allowed ($\checkmark$) or forbidden ($\times$) by the point groups of the
Sohncke groups $P2_1$, $P2_12_12_1$ and $P1$, in the frame of Fig.~\ref{fig:frame}(a).}
\begin{ruledtabular}
\begin{tabular}{llccc}
Term & Form & $C_2$ & $D_2$ & $C_1$\\
\hline
Rashba ($\alpha_1$) & $k_y\sigma_x-k_x\sigma_y$ & $\times$ & $\times$ & $\checkmark$\\
Dresselhaus ($\beta_1$) & $-(k_x\sigma_x-k_y\sigma_y)$ & $\checkmark$ & $\checkmark$ & $\checkmark$\\
Radial ($w_1$) & $k_x\sigma_x+k_y\sigma_y$ & $\checkmark$ & $\checkmark$ & $\checkmark$\\
Persistent ($\gamma_1$) & $k_x\sigma_z$ & $\checkmark$ & $\times$ & $\checkmark$\\
Persistent ($\gamma_2$) & $k_y\sigma_z$ & $\times$ & $\times$ & $\checkmark$\\
Cubic Rashba ($\alpha_3$) & see text & $\times$ & $\times$ & $\checkmark$\\
Cubic Dresselhaus ($\beta_3$) & see text & $\checkmark$ & $\checkmark$ & $\checkmark$\\
\end{tabular}
\end{ruledtabular}
\end{table}

\subsection{The persistent limit}
When $\gamma_1$ dominates, $\mathbf d=\gamma_1k_x\hat{\mathbf z}$ and $H=E_0+\hbar^2k^2/2m^*+\gamma_1k_x\sigma_z$ commutes
with $\sigma_z$. The spin along $z$ is then a good quantum number at every $\mathbf k$, with eigenvalue $s=\pm1$ of
$\sigma_z$, and completing the square gives
\begin{equation}
E_s(\mathbf k)=E_0+\frac{\hbar^2}{2m^*}\left[(k_x+s\,k_R)^2+k_y^2\right]-E_R,
\label{eq:pst}
\end{equation}
with
\begin{equation}
k_R=\frac{m^*\gamma_1}{\hbar^2},\qquad E_R=\frac{m^*\gamma_1^2}{2\hbar^2}.
\label{eq:kR}
\end{equation}
The two branches are complete paraboloids, rigidly displaced along $k_x$ by $\mp k_R$ and lowered by the Rashba energy
$E_R$, each with a uniform spin $2\langle S_z\rangle=s$ [Figs.~\ref{fig:terms}(d) and \ref{fig:cuts}(a)]. The splitting, $2|\gamma_1k_x|$, is linear in $k_x$ and vanishes
along $k_y$. Eliminating $m^*$ between $k_R$ and $E_R$ gives the model-free relation
\begin{equation}
\alpha_R\equiv\frac{2E_R}{k_R}=\gamma_1,
\label{eq:alphaR}
\end{equation}
so the coefficient can be read directly from the depth and the position of the band minimum in the calculated bands,
without a fit; deviations measure the nonparabolicity of the band.

\subsection{Persistent spin helix}
Because the spin-orbit field of every state points along $z$, a spin injected along $x$ precesses about $z$ as the electron
moves. A state with wave vector $\mathbf k$ precesses with the angular frequency $\omega=2|\mathbf d|/\hbar=2\gamma_1k_x/\hbar$
and moves along $x$ with the velocity $v_x=\hbar k_x/m_x$, where $m_x$ is the effective mass along $x$. After a distance
$x$, reached at time $t=x/v_x$, its spin has turned by
\begin{equation}
\theta(x)=\omega t=\frac{2\gamma_1k_x}{\hbar}\,\frac{m_xx}{\hbar k_x}=\frac{2m_x\gamma_1}{\hbar^2}\,x\equiv Qx.
\label{eq:theta}
\end{equation}
The angle is independent of $k_x$, so electrons with different momenta rotate their spins by the same angle over the
same distance. Momentum scattering therefore does not randomize the spin, and the spin pattern forms a persistent spin
helix~\cite{bernevig2006} with the wavelength
\begin{equation}
\lambda_{\mathrm{PSH}}=\frac{2\pi}{Q}=\frac{\pi\hbar^2}{m_x\gamma_1}.
\label{eq:psh}
\end{equation}
In practical units, $\lambda_{\mathrm{PSH}}=2\pi\,(3.81\,\mathrm{eV}\,\text{\AA}^2)/[(m_x/m_e)\gamma_1]$, using
$\hbar^2/2m_e=3.81$\,eV\,\AA$^2$; for $m_x=0.3\,m_e$ and $\gamma_1=2$\,eV\,\AA{} this gives 4\,nm.

\subsection{Enantiomers and polarization domains}\label{sec:domains}
The opposite enantiomer of a polar chiral crystal is its mirror image. The mirror $M_y$ (normal $y$, along the polar
axis) maps the S crystal onto the R crystal and at the same time reverses $\mathbf P$. A second operation, the rotation
$C_{2x}$, describes a polarization domain: it maps the $+\mathbf P$ domain of one enantiomer onto its $-\mathbf P$ domain.
The action of both operations on $\mathbf k$ and $\bm\sigma$ and on the three coefficients of Eq.~(\ref{eq:d}) is listed in
Table~\ref{tab:domains}. A mirror reverses the spin components parallel to
its plane and keeps the one along its normal, because $\bm\sigma$ is axial.

\begin{table*}[t]
\caption{\label{tab:domains}Operations relating the two enantiomers ($M_y$) and two polarization domains ($C_{2x}$) of a
polar chiral crystal: their action on
$\mathbf k$, on $\bm\sigma$, on the handedness ($\chi$) and on the polarization, and the resulting sign changes ($-$) of
the coefficients of Eq.~(\ref{eq:d}).}
\begin{ruledtabular}
\begin{tabular}{lcccccc}
Operation & $(k_x,k_y)\to$ & $(\sigma_x,\sigma_y,\sigma_z)$ & $\chi$ & $P_y$ & $\gamma_1$ & $\beta_1,w_1$\\
\hline
$M_y$ & $(k_x,-k_y)$ & $(-,+,-)$ & $-$ & $-$ & $-$ & $-$\\
$C_{2x}$ & $(k_x,-k_y)$ & $(+,-,-)$ & $+$ & $-$ & $-$ & $+$\\
\end{tabular}
\end{ruledtabular}
\end{table*}

Under $M_y$ all three coefficients change sign: the R and S enantiomers have opposite polarizations and opposite
persistent coefficients of equal magnitude, and their chirality-related coefficients $\beta_1$ and $w_1$ are opposite as
well. The persistent spin texture of the R crystal is therefore the reverse of that of the S crystal. Under $C_{2x}$ only
$\gamma_1$ changes sign, together with $\mathbf P$, so reversing the polarization of a single crystal, if it were possible,
would also reverse the spin of every conduction state.

\section{Computational methods}\label{sec:methods}
\subsection{Structures and electronic structure}
Calculations were performed with VASP~\cite{vasp1,vasp2} using the projector augmented-wave method~\cite{paw1,paw2}, the
PBE functional~\cite{pbe}, a 400\,eV cutoff and $\Gamma$-centered $4\times4\times1$ meshes. For the ten compounds the
ionic positions were relaxed at fixed lattice parameters with the D3(BJ) correction~\cite{d3,d3bj}. Opposite enantiomers were generated as exact mirror images, and the halide-substituted
analogues were fully relaxed. Polarizations were obtained from the Berry phase~\cite{ksv}. SOC was included in the non-collinear formulation~\cite{vaspsoc} with the quantization axis along
the layer normal.

\subsection{Spin textures}
Spin textures were computed non-self-consistently on a $21\times21$ grid of in-plane wave vectors with
$|k_x|,|k_y|\le0.12$\,\Ainv{} (step 0.012\,\Ainv). For each band $n$ and wave vector $\mathbf k$, VASP projects the
non-collinear wave function onto the atomic spheres and gives the charge $n^a_{n\mathbf k}$ and the magnetization
$m^a_{i,n\mathbf k}$ ($i=x,y,z$) inside the sphere of atom $a$. The spin of the state is
\begin{equation}
\langle S_i\rangle_{n\mathbf k}=\frac{\hbar}{2}\,\frac{\sum_{a\in F}m^a_{i,n\mathbf k}}{\sum_{a\in F}n^a_{n\mathbf k}},
\label{eq:spin}
\end{equation}
where the sums run over the inorganic framework $F$, i.e., Pb and the halides within 3.6\,\AA. The framework carries the SOC
and 92--99\% of the band-edge charge; dividing by the framework charge normalizes the spin so that a fully polarized state
has $|2\langle\mathbf S\rangle|=1$. We analyzed the highest valence pair and the lowest conduction pair with at least 50\%
framework weight at $\Gamma$; in \NOMBA{} this skips eight flat $\pi^*$ bands of the nitrophenyl groups 0.34\,eV below the
Pb-6$p$ conduction band. All analyzed bands satisfy the time-reversal condition
$\langle\hat{\mathbf S}(\mathbf k)\cdot\hat{\mathbf S}(-\mathbf k)\rangle=-1.00$.

\subsection{Effective masses}
Band masses were obtained from the same grids. For a Kramers pair, $E_\pm(\mathbf k)=h_0(\mathbf k)\pm|\mathbf d(\mathbf
k)|$, so the mean of the two branches is the spin-degenerate band $h_0$, whose curvature gives the mass independently of
the displacement of the band minima by the spin-orbit field. We fitted
$h_0=E_0+\hbar^2k_x^2/2m_x+\hbar^2k_y^2/2m_y+c\,k_xk_y$ within $|\mathbf k|\le0.05$\,\Ainv; the cross term vanishes by
symmetry in $P2_1$. Fit radii of 0.03 and 0.08\,\Ainv{} change the electron masses by $-2$ to $+8\%$ and the hole masses
by $-5$ to $+11\%$. Hole masses are given as positive numbers. The dispersion along the stacking direction was calculated
on a band path through $\Gamma$ with $|k_z|\le0.06$\,\Ainv.

\subsection{Model fitting and selection}
Following Ref.~\cite{ganguli2025}, for each band pair $E_0$, $m^*$ and the spin-orbit coefficients of a model were fitted to the energies of both branches
within $|\mathbf k|\le0.08$\,\Ainv{} by least squares. The 34 models (Appendix~\ref{app:models}) comprise all 15
combinations of the linear and cubic Rashba and Dresselhaus terms ($\alpha_1$, $\alpha_3$, $\beta_1$, $\beta_3$) and their
combinations with $\gamma_1$, $\gamma_2$ and $w_1$; the symmetry-forbidden terms were included to test the selection rules. Because
the energies fix only $|\mathbf d|$, each fit leaves the overall signs open; they were chosen to match the calculated spins.
The agreement of the spin directions is measured by
\begin{equation}
A=\frac{1}{2N}\sum_{\mathbf k}\sum_\pm\hat{\mathbf S}^{\mathrm{DFT}}_\pm(\mathbf k)\cdot\hat{\mathbf S}^{\mathrm{model}}_\pm(\mathbf k),
\label{eq:A}
\end{equation}
the mean cosine between the calculated and the model spin of the two branches, over the $N$ wave vectors with a splitting
of at least 2\,meV, where the spin of an energy-ordered branch is well defined. $A=1$ means identical directions at every
point, $A=0$ perpendicular spins on average. The selected model is the simplest model within 0.02 of the best $A$. Fit radii of 0.05 and 0.11\,\Ainv{} change
the conduction-band $\gamma_1$ of the strongly polar compounds by 2--5\% and select the same models (Supplemental Material). The angle disparity $\Delta\beta$
is the difference between the largest and smallest in-plane Pb-X-Pb angles~\cite{jana2021}.

\section{Results}\label{sec:results}
\subsection{Persistent texture of \NOMBA}

\begin{figure*}[t]
\includegraphics[width=\textwidth]{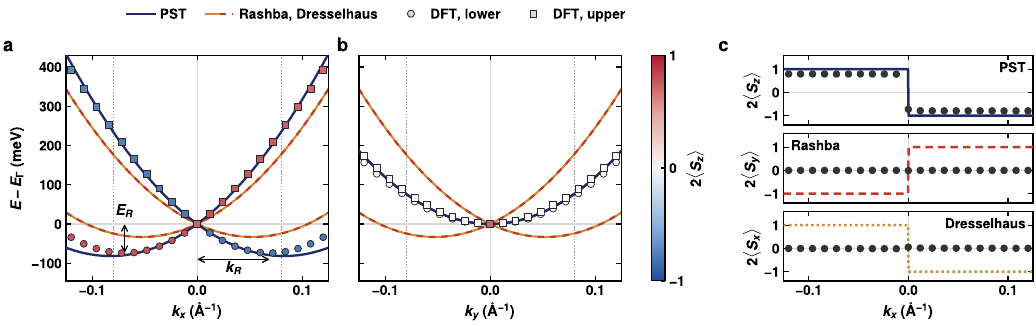}
\caption{\label{fig:cuts}Conduction-band pair of S-\NOMBA{} compared with the fitted models (PBE+SOC). (a) Cut along $k_x$ and
(b) along $k_y$ through $\Gamma$. Symbols: DFT, circles for the lower and squares for the upper branch, colored by the
framework-projected $2\langle S_z\rangle$; energies relative to the Kramers point. Navy lines: persistent model
$\gamma_1k_x\sigma_z$ ($\gamma_1=1.98$\,eV\,\AA, $m^*=0.32\,m_e$, $A=1.00$). Two-color dashed lines: best linear Rashba
and Dresselhaus fits, which have identical energies, $|\mathbf d|=\alpha|\mathbf k|$ with $\alpha=1.26$\,eV\,\AA, but
spin agreements of only 0.00 and 0.06. Dotted vertical lines: fit window $|\mathbf k|\le0.08$\,\Ainv. In (a),
$k_R=0.069$\,\Ainv{} and $E_R=74$\,meV are read directly from the DFT band, which gives the model-free
$\alpha_R=2E_R/k_R$. (c) Spin components of the lower branch along $k_x$ (DFT, symbols), each compared with the model that
predicts it: persistent $2\langle S_z\rangle=\mp1$, Rashba $2\langle S_y\rangle=\pm1$, Dresselhaus $2\langle
S_x\rangle=\mp1$ for $k_x\gtrless0$. The calculated spins point out of the layer ($|2\langle S_z\rangle|\approx0.8$), and
their in-plane components vanish.}
\end{figure*}

The compound with the largest polarization, S-\NOMBA{} ($P_y=3.85$\,\uC), shows the signature of
Eq.~(\ref{eq:pst}) directly in its bands (Fig.~\ref{fig:cuts}). Along $k_x$ the conduction-band pair consists of two
parabolas that cross at $\Gamma$ and reach their minima at $k_x=\pm0.069$\,\Ainv, and each parabola carries a single sign
of $S_z$. Along $k_y$ the branches stay within 17\,meV of each other, compared with a splitting of 240\,meV at
$k_x=0.06$\,\Ainv, and the in-plane spin stays below 0.09 while $|2\langle S_z\rangle|\approx0.8$. The persistent model,
with $\gamma_1=1.98$\,eV\,\AA{} and $m^*=0.32\,m_e$, follows both cuts inside the fit window; outside it the lower branch
flattens because of nonparabolicity. The Rashba-only and Dresselhaus-only fits give identical energies, because both have
$|\mathbf d|=\alpha|\mathbf k|$; they cannot follow the anisotropy of the splitting (energy error 34\,meV), and
they place the spins in the layer [Fig.~\ref{fig:cuts}(c)].

\begin{figure*}[t]
\includegraphics[width=\textwidth]{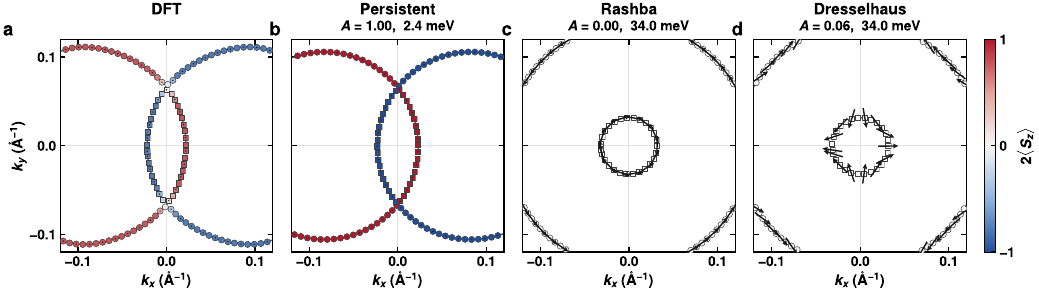}
\caption{\label{fig:models}Constant-energy contours of the conduction-band pair of S-\NOMBA{} 52\,meV above the Kramers point.
(a) DFT (PBE+SOC); (b) persistent model $\gamma_1k_x\sigma_z$; (c) linear Rashba model $\alpha_1(k_y\sigma_x-k_x\sigma_y)$;
(d) linear Dresselhaus model $-\beta_1(k_x\sigma_x-k_y\sigma_y)$. Lower branch: circles and solid contours; upper branch:
squares and dashed contours; color: $2\langle S_z\rangle$; arrows: in-plane spin. Each model panel gives the spin agreement
$A$ [Eq.~(\ref{eq:A})] and the energy RMSE of its fit to both branches within $|\mathbf k|\le0.08$\,\Ainv. Only the
persistent model reproduces the two displaced circles with uniform out-of-plane spin ($A=1.00$, 2.4\,meV), against
$A=0.00$ and 0.06 with 34\,meV for the Rashba and Dresselhaus models.}
\end{figure*}

The constant-energy contours show the same result (Fig.~\ref{fig:models}). The calculated contours are two
circles displaced along $\pm k_x$, each with a uniform out-of-plane spin. The persistent model reproduces them with $A=1.00$
and an energy error of 2.4\,meV. The best linear Rashba and Dresselhaus models fit the energies with an error of 34\,meV
but place the spins in the layer, perpendicular to the calculated spins ($A=0.00$ and 0.06). The model-free parameter of
the DFT band, $\alpha_R=2.15$\,eV\,\AA{} ($E_R=74$\,meV, $k_R=0.069$\,\Ainv), exceeds $\gamma_1$ by 9\% because of the
nonparabolicity. The valence-band pair is also persistent, with $\gamma_1=-0.50$\,eV\,\AA{} and $A=1.00$.

\subsection{Model selection across the family}

\begin{figure*}[t]
\includegraphics[width=\textwidth]{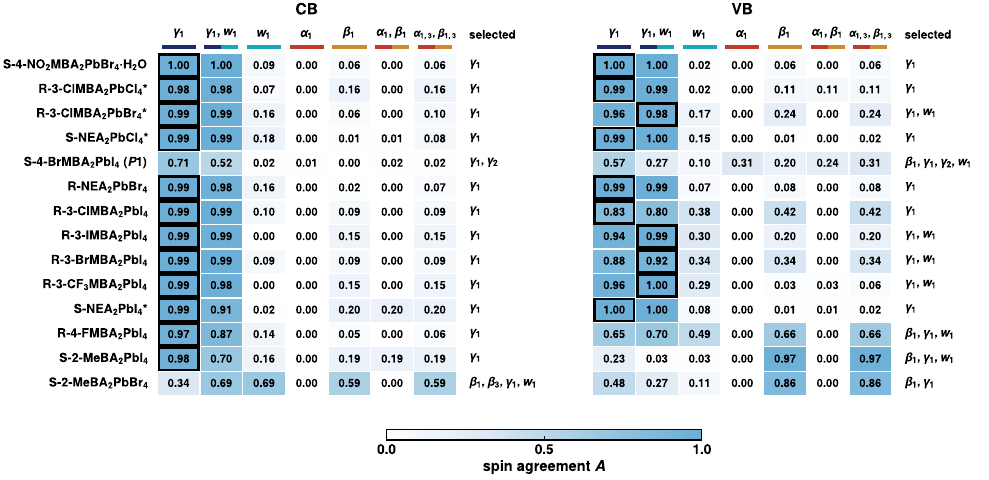}
\caption{\label{fig:selection}Model selection by spin agreement. For each polar compound (one enantiomer; its mirror image gives the
same $A$) and for the conduction (CB, left) and valence (VB, right) band pair, the spin agreement $A$ [Eq.~(\ref{eq:A})] of
the main models fitted to the same DFT bands: persistent $\gamma_1$; persistent plus radial ($\gamma_1,w_1$); radial $w_1$;
linear Rashba $\alpha_1$; linear Dresselhaus $\beta_1$; linear Rashba plus Dresselhaus ($\alpha_1,\beta_1$); and the best
model built only from linear and cubic Rashba and Dresselhaus terms ($\alpha_{1,3},\beta_{1,3}$). The colored bars under the
column headers mark the terms each model contains (persistent navy, Rashba red, Dresselhaus amber, radial teal). A black
frame marks the selected model, the simplest within 0.02 of the best $A$; its terms are listed at the right, including terms
outside the shown columns (e.g., $\gamma_1,\gamma_2$ for the $P1$ host). Asterisks: halide-substituted analogues.}
\end{figure*}

\begin{table*}[t]
\caption{\label{tab:fits}Fitted spin-orbit fields of the conduction- (CB) and valence-band (VB) pairs. $P_y$: polarization
of the computed enantiomer along the $2_1$ axis (\textmu C\,cm$^{-2}$); $\Delta\beta$: disparity of the in-plane Pb-X-Pb
angles (deg). Model: terms of the selected model [Eq.~(\ref{eq:d}); $\beta_3$ is the cubic Dresselhaus term]. $\gamma_1$ in
eV\,\AA; $\alpha_R=2E_R/k_R$ from the DFT band (-- where no off-$\Gamma$ minimum is resolved); $A$: spin agreement of the
selected model; $A_{\mathrm{RD}}$: best model built from Rashba and Dresselhaus terms only. Mirror enantiomers have the same
values with opposite signs of $P_y$ and of all coefficients. Asterisks mark the halide-substituted analogues.}
\begin{ruledtabular}
\begin{tabular}{lrrlrrrrlrr}
 & & & \multicolumn{5}{c}{CB} & \multicolumn{3}{c}{VB}\\
\cline{4-8}\cline{9-11}
Compound & $P_y$ & $\Delta\beta$ & Model & $\gamma_1$ & $\alpha_R$ & $A$ & $A_{\mathrm{RD}}$ & Model & $\gamma_1$ & $A$\\
\hline
S-4-NO$_{2}$MBA$_{2}$PbBr$_{4}$$\cdot$H$_{2}$O & +3.85 & 20.6 & $\gamma_1$ & +1.98 & 2.15 & 1.00 & 0.06 & $\gamma_1$ & $-$0.50 & 1.00 \\
R-3-ClMBA$_{2}$PbCl$_{4}$$^{\ast}$ & $-$2.96 & 20.5 & $\gamma_1$ & $-$1.33 & 1.40 & 0.98 & 0.16 & $\gamma_1$ & +0.14 & 0.99 \\
R-3-ClMBA$_{2}$PbBr$_{4}$$^{\ast}$ & $-$2.60 & 16.4 & $\gamma_1$ & $-$1.23 & 1.30 & 0.99 & 0.10 & $\gamma_1,w_1$ & +0.18 & 0.98 \\
S-NEA$_{2}$PbCl$_{4}$$^{\ast}$ & +2.57 & 25.5 & $\gamma_1$ & +1.64 & 1.75 & 0.99 & 0.08 & $\gamma_1$ & $-$0.09 & 0.99 \\
S-4-BrMBA$_{2}$PbI$_{4}$\footnotemark[1] & $-$2.52 & 7.1 & $\gamma_1,\gamma_2$ & +0.49\,/\,$-$0.30 & 0.66 & 1.00 & 0.01 & $\beta_1,\gamma_1,\gamma_2,w_1$ & $-$0.23\,/\,+0.21 & 0.96 \\
R-NEA$_{2}$PbBr$_{4}$ & $-$2.49 & 17.2 & $\gamma_1$ & $-$1.35 & 1.42 & 0.99 & 0.07 & $\gamma_1$ & +0.26 & 0.99 \\
R-3-ClMBA$_{2}$PbI$_{4}$ & $-$2.43 & 13.5 & $\gamma_1$ & $-$1.09 & 1.16 & 0.99 & 0.09 & $\gamma_1$ & +0.13 & 0.83 \\
R-3-IMBA$_{2}$PbI$_{4}$ & $-$2.42 & 14.1 & $\gamma_1$ & $-$1.20 & 1.28 & 0.99 & 0.15 & $\gamma_1,w_1$ & +0.21 & 0.99 \\
R-3-BrMBA$_{2}$PbI$_{4}$ & $-$2.40 & 14.2 & $\gamma_1$ & $-$1.17 & 1.25 & 0.99 & 0.09 & $\gamma_1,w_1$ & +0.14 & 0.92 \\
R-3-CF$_{3}$MBA$_{2}$PbI$_{4}$ & $-$2.37 & 14.8 & $\gamma_1$ & $-$1.28 & 1.38 & 0.99 & 0.15 & $\gamma_1,w_1$ & +0.22 & 1.00 \\
S-NEA$_{2}$PbI$_{4}$$^{\ast}$ & +2.34 & 13.9 & $\gamma_1$ & +1.37 & 1.45 & 0.99 & 0.20 & $\gamma_1$ & $-$0.27 & 1.00 \\
R-4-FMBA$_{2}$PbI$_{4}$ & +1.13 & 0.7 & $\gamma_1$ & +0.03 & -- & 0.97 & 0.05 & $\beta_1,\gamma_1,w_1$ & $-$0.02 & 0.95 \\
S-2-MeBA$_{2}$PbI$_{4}$ & +0.01 & 0.8 & $\gamma_1$ & $-$0.14 & -- & 0.98 & 0.19 & $\beta_1,\gamma_1,w_1$ & +0.02 & 1.00 \\
S-2-MeBA$_{2}$PbBr$_{4}$ & 0.00 & 0.0 & $\beta_1,\beta_3,\gamma_1,w_1$ & +0.02 & -- & 0.92 & 0.59 & $\beta_1,\gamma_1$ & 0.00 & 0.97 \\
\end{tabular}
\end{ruledtabular}
\footnotetext[1]{Space group $P1$: the polar axis is not fixed by symmetry; the second value is $\gamma_2$ of $k_y\sigma_z$.}
\end{table*}

Figure~\ref{fig:selection} compares the spin agreement of the main model classes across the family, and
Table~\ref{tab:fits} lists the selected models. In the six strongly polar pristine compounds
($|P_y|=2.37$--3.85\,\uC) and the four substituted analogues, the conduction band is described by $\gamma_1k_x\sigma_z$
alone, with $|\gamma_1|=1.09$--1.98\,eV\,\AA{} and $A\ge0.98$. The best model built from Rashba and Dresselhaus terms never
exceeds $A=0.20$, although its energy error is often small. Adding the radial or Dresselhaus terms allowed by $C_2$ does
not improve the persistent model beyond the 0.02 threshold, and the symmetry-forbidden terms $\alpha_1$, $\alpha_3$ and
$\gamma_2$ never improve it: the calculations reproduce the selection rules of Table~\ref{tab:sym}.

\subsection{Magnitude of $\gamma_1$ and the framework distortion}

\begin{figure*}[t]
\includegraphics[width=\textwidth]{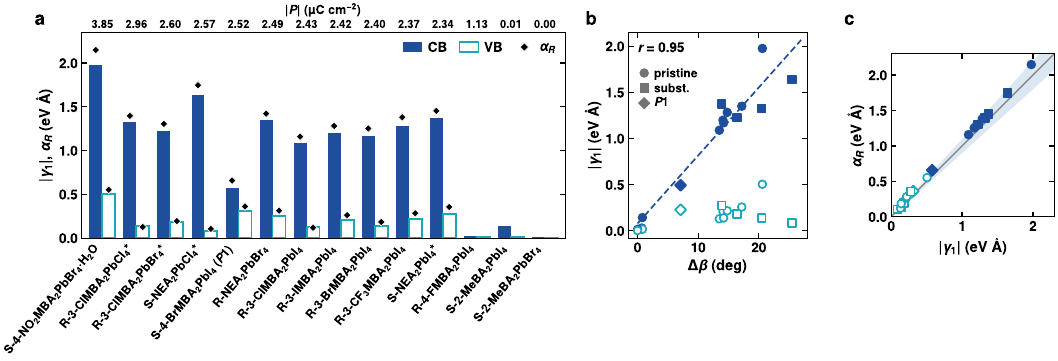}
\caption{\label{fig:gamma}The persistent coefficient across the family. (a) $|\gamma_1|$ of the selected model for the conduction
(filled) and valence (open) band, ordered by $|P_y|$ (top axis, \uC); diamonds: model-free $\alpha_R=2E_R/k_R$ read from the
DFT bands; for the $P1$ host the bar gives $(\gamma_1^2+\gamma_2^2)^{1/2}$. (b) $|\gamma_1|$ against the disparity
$\Delta\beta$ of the in-plane Pb-X-Pb angles; dashed line: linear fit to the conduction-band values, with Pearson
coefficient $r=0.95$; circles, squares and diamonds: pristine, halide-substituted and $P1$ compounds. (c) $\alpha_R$ against
$|\gamma_1|$; line: $\alpha_R=|\gamma_1|$ [Eq.~(\ref{eq:alphaR})]; shaded band: $\pm10\%$.}
\end{figure*}

The magnitude of $\gamma_1$ is set by the inorganic framework (Fig.~\ref{fig:gamma}). $|\gamma_1^{\mathrm{CB}}|$ grows linearly with $\Delta\beta$,
$|\gamma_1|\approx0.074\,\Delta\beta+0.08$\,eV\,\AA{} with $\Delta\beta$ in degrees ($r=0.95$), and correlates more closely
with $\Delta\beta$ than with $|P_y|$ ($r=0.88$). R-4-FMBA$_2$PbI$_4$, with $|P_y|=1.13$\,\uC{} but
$\Delta\beta=0.7^\circ$ and $\gamma_1\approx0$, illustrates the difference. The conduction states are Pb-6$p$ states, and
their splitting measures the inversion asymmetry of the Pb-X network in which they reside, whereas the total polarization
also contains dipoles of the organic cations. The halide series follows the same rule: in NEA$_2$PbX$_4$ the chloride has
the largest splitting ($\gamma_1=1.64$\,eV\,\AA) despite the lighter halide, because its framework is the most distorted
($\Delta\beta=25.5^\circ$). These results extend the $\Delta\beta$ descriptor of Jana \textit{et al.}~\cite{jana2021},
proposed for Rashba-Dresselhaus splittings, to persistent spin textures. The model-free $\alpha_R$ of Eq.~(\ref{eq:alphaR})
exceeds the fitted $\gamma_1$ by 5--9\% in every strongly polar compound [Fig.~\ref{fig:gamma}(c)], as expected from the
nonparabolicity of the band, so it is a useful first estimate of the persistent coefficient.

\subsection{Opposite persistent spin textures of the two enantiomers}

\begin{figure*}[t]
\includegraphics[width=0.68\textwidth]{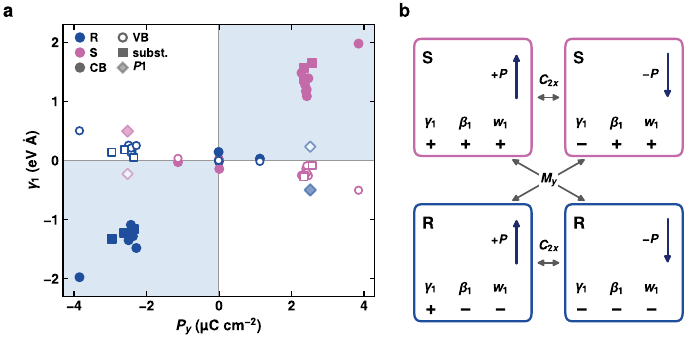}
\caption{\label{fig:sign}Sign of the persistent coefficient in both enantiomers. (a) Signed $\gamma_1$ of the
conduction (filled) and valence (open) band against the polarization $P_y$ of the computed structure, for both enantiomers
(R navy, S orchid) of every compound, with the substituted analogues as squares. In the shaded quadrants $\gamma_1$ has the
sign of $P_y$. In all 20 $P2_1$ structures with $|P_y|>2$\,\uC, $\gamma_1^{\mathrm{CB}}$ has the sign of $P_y$ (20/20) and
$\gamma_1^{\mathrm{VB}}$ the opposite sign (20/20). The $P1$ host (faded diamonds) has no twofold axis and is not bound by
the rule. (b) Enantiomers and polarization domains of one compound, with the signs of $\gamma_1$, $\beta_1$ and $w_1$ relative to
(S,\,$+\mathbf P$) from Table~\ref{tab:domains}: the mirror $M_y$ relates the S and R crystals and reverses $\mathbf P$ and
all three coefficients; $C_{2x}$ relates two polarization domains and reverses $\mathbf P$ and $\gamma_1$.}
\end{figure*}

Both enantiomers of every compound were computed, which gives 20 $P2_1$ structures with $|P_y|>2$\,\uC. In all of them
$\gamma_1$ of the conduction band has the sign of $P_y$, whether the structure is R or S (Fig.~\ref{fig:sign}), and $\gamma_1$
of the valence band has the opposite sign; the conduction and valence pairs therefore carry opposite persistent spins for
the same $k_x$. In every compound the two enantiomers have opposite polarizations and give the same $|\gamma_1|$ with opposite sign
(Table S6), as required by the mirror $M_y$ of Table~\ref{tab:domains}. This holds also for NEA$_2$PbBr$_4$, whose R and S
crystal structures were determined independently: $\gamma_1=-1.350$ and $+1.389$\,eV\,\AA, and $\pm1.48$\,eV\,\AA{} when
both are relaxed in the same cell. The R and S crystals of each compound therefore carry opposite persistent spin
textures.

\subsection{Valence bands, low symmetry and weak distortion}
The valence bands are also persistent but weaker ($|\gamma_1|=0.09$--0.50\,eV\,\AA; persistent-model agreement
0.83--1.00, against at most 0.42 for any Rashba or Dresselhaus model). They are formed mainly by halide
$p$ states with a smaller atomic SOC than Pb-6$p$, and in four of them a small radial term ($|w_1|\le0.07$\,eV\,\AA)
improves the agreement. The triclinic host crystal S-4-BrMBA$_2$PbI$_4$ ($P1$) has no twofold axis, so its polarization is
not tied to $y$. Both $k_x\sigma_z$ and $k_y\sigma_z$ appear ($\gamma_1=0.49$, $\gamma_2=-0.30$\,eV\,\AA, $A=1.00$): the spins
remain normal to the layer, but the splitting direction is rotated by $31^\circ$ from $x$, i.e., the persistent field
follows the actual in-plane polar direction. R-4-FMBA$_2$PbI$_4$ and the two 2-MeBA compounds have nearly undistorted
frameworks ($\Delta\beta\le0.8^\circ$). Their splittings are small ($|\gamma_1|\le0.14$\,eV\,\AA), no off-$\Gamma$ minimum is
resolved, and four of their six selected models add Dresselhaus or radial terms to the persistent one (Table~\ref{tab:fits}),
so their classification is less certain.

\subsection{Effective masses}\label{sec:mass}

\begin{table}[h]
\caption{\label{tab:mass}In-plane effective masses ($m_e$) of the conduction-band (electron) and valence-band (hole) pairs
from the PBE+SOC bands, along $x$ and along the $2_1$ axis $y$. The two enantiomers have identical masses. Asterisks mark
the halide-substituted analogues.}
\begin{ruledtabular}
\begin{tabular}{lcccc}
 & \multicolumn{2}{c}{Electron} & \multicolumn{2}{c}{Hole}\\
\cline{2-3}\cline{4-5}
Compound & $m_x$ & $m_y$ & $m_x$ & $m_y$\\
\hline
4-NO$_{2}$MBA$_{2}$PbBr$_{4}$$\cdot$H$_{2}$O & 0.29 & 0.32 & 0.54 & 0.45 \\
3-ClMBA$_{2}$PbCl$_{4}$$^{\ast}$ & 0.41 & 0.48 & 0.54 & 0.64 \\
3-ClMBA$_{2}$PbBr$_{4}$$^{\ast}$ & 0.32 & 0.36 & 0.48 & 0.51 \\
NEA$_{2}$PbCl$_{4}$$^{\ast}$ & 0.43 & 0.51 & 1.04 & 1.09 \\
4-BrMBA$_{2}$PbI$_{4}$ & 0.23 & 0.22 & 0.45 & 0.45 \\
NEA$_{2}$PbBr$_{4}$ & 0.33 & 0.37 & 0.55 & 0.57 \\
3-ClMBA$_{2}$PbI$_{4}$ & 0.27 & 0.29 & 0.62 & 0.47 \\
3-IMBA$_{2}$PbI$_{4}$ & 0.26 & 0.28 & 0.53 & 0.46 \\
3-BrMBA$_{2}$PbI$_{4}$ & 0.27 & 0.28 & 0.57 & 0.46 \\
3-CF$_{3}$MBA$_{2}$PbI$_{4}$ & 0.26 & 0.29 & 0.61 & 0.49 \\
NEA$_{2}$PbI$_{4}$$^{\ast}$ & 0.30 & 0.32 & 0.56 & 0.46 \\
4-FMBA$_{2}$PbI$_{4}$ & 0.22 & 0.21 & 0.42 & 0.46 \\
2-MeBA$_{2}$PbI$_{4}$ & 0.20 & 0.20 & 0.30 & 0.30 \\
2-MeBA$_{2}$PbBr$_{4}$ & 0.26 & 0.26 & 0.31 & 0.31 \\
\end{tabular}
\end{ruledtabular}
\end{table}

The band masses are listed in Table~\ref{tab:mass}. The conduction bands are light in the layer, with
$m_x=0.20$--0.43\,$m_e$ and $m_y=0.20$--0.51\,$m_e$, and flat along the stacking direction: no conduction pair disperses
by more than 0.2\,meV within $|k_z|\le0.06$\,\Ainv{} ($m_z\gtrsim50\,m_e$), and the valence pairs disperse by at most
1.2\,meV, except in NEA$_2$PbCl$_4$ (6\,meV). The holes are 1.2--2.4 times heavier than the electrons. In the ten strongly
polar $P2_1$ compounds the electron mass along the polar axis exceeds that along $x$ by 7--17\%, whereas the three weakly
distorted compounds and the $P1$ host are isotropic to within 4\%. The electron mass increases from the iodides to the
chlorides, from 0.27 to 0.41\,$m_e$ along $x$ in 3-ClMBA$_2$PbX$_4$ and from 0.30 to 0.43\,$m_e$ in NEA$_2$PbX$_4$.

The kinetic term of Eq.~(\ref{eq:H}) does not act on the spin, so the mass anisotropy does not affect the persistent
texture. With $\hbar^2k_x^2/2m_x+\hbar^2k_y^2/2m_y$ in place of $\hbar^2k^2/2m^*$, $H$ still commutes with $\sigma_z$, and
Eq.~(\ref{eq:kR}) holds with $m^*$ replaced by $m_x$. With the DFT $m_x$ and the fitted $\gamma_1$, it gives
$E_R=75$\,meV for \NOMBA, against 74\,meV read from the band, and reproduces $E_R$ to within 6\% in all strongly polar
compounds; $k_R$ is overestimated by up to 12\% because the band flattens beyond the fit radius. The isotropic $m^*$ of the
fitted models (Supplemental Material) is 3--8\% larger than the harmonic mean of $m_x$ and $m_y$, since it is fitted over
the larger radius of 0.08\,\Ainv.

\subsection{Persistent spin helices}

\begin{figure*}[t]
\includegraphics[width=0.92\textwidth]{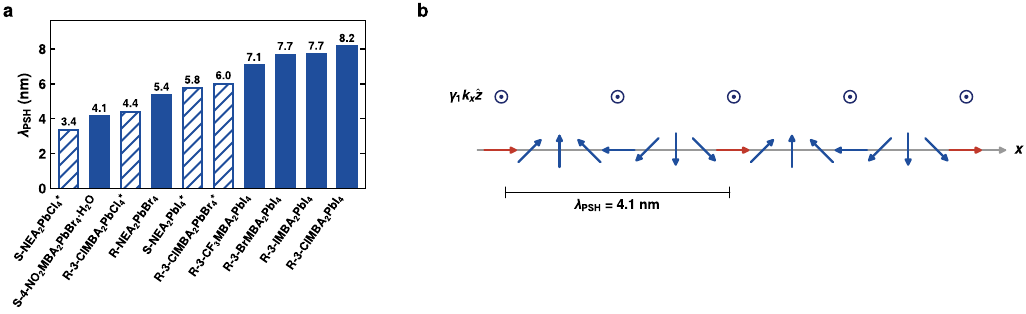}
\caption{\label{fig:psh}Persistent spin helix. (a) Wavelength $\lambda_{\mathrm{PSH}}=\pi\hbar^2/(m_x\gamma_1)$
[Eq.~(\ref{eq:psh})] of the conduction-band helix of the strongly polar $P2_1$ compounds, with $m_x$ the DFT band mass along
$x$; hatched bars: substituted analogues. (b) Schematic for the conduction band of S-\NOMBA: an electron moving along $+x$
feels the spin-orbit field $\gamma_1k_x\hat{\mathbf z}$, normal to the layer ($\odot$); a spin injected along $x$ (red)
rotates in the layer plane with the wave vector $Q=2m_x\gamma_1/\hbar^2$, independent of its momentum, and returns to its
initial direction after $\lambda_{\mathrm{PSH}}=4.1$\,nm.}
\end{figure*}

With $m_x=0.29\,m_e$ (Table~\ref{tab:mass}) and $\gamma_1=1.98$\,eV\,\AA, Eq.~(\ref{eq:psh}) gives
$\lambda_{\mathrm{PSH}}=4.1$\,nm for the conduction band of \NOMBA, and the strongly polar compounds span 3.4--8.2\,nm
(Fig.~\ref{fig:psh}). The wavelength depends on the product $m_x\gamma_1$: the shortest helix, 3.4\,nm, is found in
NEA$_2$PbCl$_4$, whose $\gamma_1$ (1.64\,eV\,\AA) is smaller than that of \NOMBA{} but whose electron mass along $x$ is
1.5 times larger. These wavelengths are
three orders of magnitude shorter than the micrometer-scale helices observed in GaAs quantum wells~\cite{koralek2009},
where the helix requires tuned Rashba and Dresselhaus couplings that are correspondingly weaker. A helix with a period of a
few nanometers is protected against momentum scattering in the same way, but its observation requires a single polarization
domain, because the persistent direction reverses with $\mathbf P$.

\section{Discussion}
The persistent texture found here belongs to the class of symmetry-enforced PSTs identified by Tao and
Tsymbal~\cite{tao2018}: an in-plane polar axis turns the Rashba field into a unidirectional field normal to the layer. The
chiral polar perovskites differ from the achiral polar systems studied before in two respects. First, their point group allows,
in addition, chirality-related Dresselhaus and radial terms; the calculations show that these terms are small in the
conduction band, so the texture remains persistent to within 2\% in the spin directions. Second, the chiral crystals come as two
enantiomers with opposite polarizations, and the persistent spin texture reverses between them. Chirality thus enters at
two levels: the chiral cations crystallize the layers in the Sohncke group $P2_1$, whose polar $2_1$ axis lies in the layer
and enforces the persistent texture, and the two enantiomers carry persistent spin textures of opposite sign.

The analysis also shows why band energies are insufficient to classify spin-orbit coupling in low-symmetry crystals. All
linear terms give the same $|\mathbf d|$ and hence the same bands, and models with entirely wrong spin directions fit the
energies with errors of a few tens of meV. A classification based on the band splitting alone, such as the Rashba
parameter $\alpha_R$, cannot distinguish a helical Rashba texture from a persistent one, whereas the spin directions do.

Two limitations should be noted. First, the two-band model describes each Kramers pair separately and is valid near $\Gamma$; at
larger $|\mathbf k|$ the bands become nonparabolic, which is why $\alpha_R$ exceeds $\gamma_1$ slightly. Second, the
calculations use PBE, which underestimates band gaps but describes the SOC splitting of the Pb-6$p$ conduction band, which
is an intra-band property, reliably. Spin-resolved photoemission of the conduction band in doped samples or optical
spin-grating measurements of the helix wavelength would test the predictions directly. Whether the persistent spin
direction can be reversed in practice depends on the barrier for reversing $\mathbf P$.

\section{Conclusions}
In chiral 2D perovskites with space group $P2_1$, the in-plane polar axis forbids the helical Rashba field and converts the
Rashba interaction into a persistent field along the layer normal. Spin textures of fourteen compounds obtained from first principles calculations,
analyzed with 34 two-band models ranked by their spin directions, confirm the following: the conduction bands are described by
$\gamma_1k_x\sigma_z$ alone, with $\gamma_1$ up to 2\,eV\,\AA{} and helix wavelengths of a few nanometers. The conduction
electrons are light in the layer (0.20--0.43\,$m_e$ along $x$) and do not disperse along the stacking direction, and the
helix wavelength is set by the product of $\gamma_1$ and this mass. The R and S
enantiomers have opposite polarizations, and therefore opposite $\gamma_1$ of equal magnitude and opposite persistent spin
textures, as found for all ten compounds. Therefore, the choice of enantiomer selects the spin direction. The magnitude of $\gamma_1$ is set by the angle disparity of the
framework, which gives a structural route to large persistent splittings.

\begin{acknowledgments}
P.G. and J.F. acknowledge Olle Engkvists Stiftelse for financial support.
P.G. acknowledges financial support from the strength area \emph{AI for research} at Uppsala University.
B.S. acknowledges financial support from Swedish Research Council (grant no.~2022-04309), STINT Mobility Grant for
Internationalization (grant no.~MG2022-9386) and Carl Tryggers Stiftelse (grant no.~25:4513).
Computational resources were provided by the National Academic Infrastructure for Supercomputing in Sweden (NAISS),
funded by the Swedish Research Council.
The calculations were performed under projects NAISS 2025/22-1414, UPPMAX 2026/1-32 and NAISS 2026/1-46.
\end{acknowledgments}

\section*{Data availability}
The data that support the findings of this article are available from the authors upon reasonable request.

\appendix
\section{The 34 fitted models}\label{app:models}
Table~\ref{tab:models} lists the models. The first 15 contain only the Rashba and Dresselhaus terms used in many
analyses of perovskite spin textures; the remaining 19 add the terms allowed by chiral and in-plane polar point groups.
Symmetry-forbidden terms (for $C_2$: $\alpha_1$, $\alpha_3$, $\gamma_2$) are included as a test of the selection rules.

\begin{table}[h]
\caption{\label{tab:models}The 34 two-band models: R1, R3, D1, D3, PST, PSTy and W1 denote the terms with coefficients
$\alpha_1$, $\alpha_3$, $\beta_1$, $\beta_3$, $\gamma_1$, $\gamma_2$ and $w_1$.}
\begin{ruledtabular}
\begin{tabular}{p{2.2cm}p{5.6cm}}
Group & Models\\
\hline
Rashba, Dresselhaus (15) & every non-empty combination of R1, R3, D1, D3\\
Radial (5) & W1, R1+W1, D1+W1, R1+D1+W1, D1+D3+W1\\
Persistent (13) & PST, PST+W1, PST+PSTy, PST+PSTy+W1, D1+PST, R1+PST, R1+D1+PST, D1+D3+PST, D1+PST+W1, R1+PST+W1,
R1+D1+PST+W1, D1+PST+PSTy+W1, D1+D3+PST+W1\\
All terms (1) & R1+R3+D1+D3+PST+PSTy+W1\\
\end{tabular}
\end{ruledtabular}
\end{table}

\bibliography{references}

\end{document}